\documentclass[preprint,aps,prl,showpacs,superscriptaddress]{revtex4}
\usepackage{graphics}
\usepackage{bm}
\begin{document}
\title{Electric Conduction in Short DNA Wires}
\author{Weihua \surname{MU}}
\email[Email address: ] {muwh@itp.ac.cn} \affiliation{Institute of
Theoretical Physics,
 The Chinese Academy of Sciences,
 P.O.Box 2735 Beijing 100080, China}
\author{Zhong-can \surname{Ou-Yang}}
\affiliation{Institute of Theoretical Physics,
 The Chinese Academy of Sciences,
 P.O.Box 2735 Beijing 100080, China}
\affiliation{Center for Advanced Study,
 Tsinghua University, Beijing 100084, China}
\begin{abstract}
%%%%%%%%%%%%%%%%%%%%%%%%%%%%%%%%%%%%%
A strict method is used to calculate the current-voltage
characteristics of a double-stranded DNA. A more reliable model
considering the electrostatic potential drop along an individual
DNA molecular wire between the contacts is considered and the
corresponding Green's Function is obtained analytically using
Generating Function method, which avoids difficult numerical
evaluations. The obtained results indicate that the electrostatic
drop along the wire always increases the conductor beyond the
threshold than without considering it, which is in agreement with
recent experiments. The present method can also be used to
calculate the current-voltage characteristics for other molecular
wires of arbitrary length.
\end{abstract}
 \maketitle
%%%%%%%%%%%%%%%%%%%%%%%%%%%%%%%%%%%%%%%%%%%%%%%%%%%%%%%%%
Recently, the progress of molecular wires \cite{A} attracts much
attention on their transport behaviors. A number of experimental
groups have reported measurements of the current-voltage (I-V)
characteristics of molecules, especially Deoxyribonucleic Acid
(DNA)\cite{AR,KH}. DNA has a special double-helix structure with
complementary nucleotide base-pairs stacking onto each other,
which could possibly be a candidate for one-dimensional electronic
transport\cite{10}. Intense experimental investigations have
already been made on the transport properties of DNA\cite{C}.
Using scanning tunneling microscope technique, Dunlap {\it et al}.
\cite{D} found that DNA is an insulator. Fink and schoenenberger
\cite{F} measured I-V characteristics of $\lambda$-DNA ropes
consisting of a few double-stranded DNA (dsDNA) molecules by low
energy electron bombardment technique, and found a linear
current-voltage relation. Kasumov {\it et al}. \cite{K} measured
small number of DNA molecules, and observed proximity-induced
superconductivity. In particular, Porath {\it et al}. \cite{P}
directly measured electrical transport through individual DNA
molecules, and suggested DNA to be a semiconductor with a voltage
gap. To explain experimental disputes, several theoretical models
ranging from the H\"{u}kel model \cite{6} to the density
functional theory \cite {den,Mattias} have been developed.

In the present paper, we investigate theoretically Porath's
experiment \cite{P} and calculate the current between two
electrodes which are connected by a DNA molecular wire. To use a
simple model illustrating the basic physics, we follow the
algorithm proposed by Mujica {\it et al}. \cite{6}. The model
Hamiltonian can be written as the sum of two terms,
\begin{equation}
H=H^{0}+H^{'},
\end{equation}
where $H^{0}$ is the non-interacting Hamiltonian for the
electrodes (electron reservoirs) and DNA molecule, and $H^{'}$
represents the coupling of contacts and DNA molecule. Using
extended H\"{u}kel model, $H^{0}$ and $H^{'}$ can be expressed as
\cite{6}
\begin{eqnarray}
  H^{0} &=&\sum_{i}E_{i}^0\mid i\rangle\langle i\mid+\sum_{\alpha=1}^N
 E_{\alpha}^0\mid \alpha\rangle\langle\alpha\mid+\sum_{f}E_{f}^0\mid f\rangle\langle f
 \mid + \\ \nonumber
 && \sum_{\alpha=1}^{N-1}\beta\mid\alpha\rangle\langle{\alpha+1}\mid+h.c.
 \\
 H^{'}&=&\sum_{i}V_{i1}\mid i\rangle\langle1\mid +\sum_{f}V_{Nf}\mid
 N\rangle\langle f\mid+h.c. ,\\ \nonumber
\end{eqnarray}
where  the sum on $\mid i\rangle$ ($\mid f\rangle$) runs over the
state in the continuum of left (right) reservoirs. A DNA molecular
wire consists of N sites, with one state per site, which is
denoted by $\mid\alpha \rangle$. $V_{i1}$ ( $V_{Nf}$) is the
tunneling parameter from the left (right) reservoir to the single
electron state $1(N)$ in the molecule. T-matrix formalism of
scattering theory \cite{T} gives differential conductance
\cite{6}:
\begin{equation} g=\frac{2e^{2}}{\pi\hbar}\mid G_{1N}\mid
 ^2\Delta_{A}(E_{f})\Delta_{B}(E_{f}),
 \end{equation}
where $ G_{1N} $ is the $(1N)$ element of the matrix formula
Green's Function G, it is related to the molecule's site $1$ and
$N$. $\Delta_{1}$ and $\Delta_{N}$ are semielliptical reservoir
spectral densities defined through Newns' chemisorption theory
\cite{N}:
\begin{equation}
\Delta_{K}(E)=\left\{
\begin{array}{cc}
\frac{V_{K}^2}{\gamma}\sqrt{1-(E/2\gamma )^2} ,& \hspace{2cm} \mid
E/2\gamma \mid <1,\\
0,& \hspace{2.0cm}\mid E/2\gamma \mid >1,
\end{array}\right.
\end{equation}
where E is measured from the center of reservoir energy band
caused by chemisorption in the surface of the electrode. $V_{K}$
$(K=1, N)$ describe the strength of the chemisorption coupling
between the DNA molecule and the reservoirs, and $4\gamma$ is the
bandwidth of the reservoir. Through L\"{o}wdin's matrix partition
technique \cite{15}, the electrode Hamiltonian can be replaced by
a self-energy:
\begin{eqnarray}
H_{eff}&=& \Sigma_{1}\mid 1\rangle\langle 1\mid+\sum_{\alpha=1}^N
 E_{\alpha}^0\mid
 \alpha\rangle\langle\alpha\mid+\sum_{\alpha=1}^{N-1}\beta\mid\alpha\rangle\langle{\alpha+1}\mid
 \\ \nonumber
 && +\Sigma_{N} \mid N\rangle\langle N \mid+h.c.,
\end{eqnarray}
where $\Sigma_{K}$ $(K=1, N)$ are respectively the self-energy
resulting from the coupling of the molecule to the left (right)
electrode. The Green's Function is expressed as
\begin{equation}
 G=\frac{1} {(zI-H_{eff})},
 \end{equation}
where z is a complex number,  whose real part $E$ is the energy of
the transfer electrons.

Mujica {\it et al}. \cite{6} ignored the electrostatic drop along
the molecule. Here we consider the linear voltage drop along the
molecule, i.e., the electric potential between the electrodes
varying linearly with distance. This assumption is a good
approximation to the computed potential profile through the
molecule between the two electrodes \cite{RE}. Thus the energies
of the sites are function of bias V:
\begin{eqnarray}
E_{\alpha}^{0}&=& E_{b}-qV_{0}(\alpha -1), \;\;\;\;\; \alpha=(1,...,N),\\
         V_{0}&=& V/(N-1),
\end{eqnarray}
The effective Hamiltonian can be expressed in the following matrix
form:
\begin{equation}
H_{eff}= \left[
\begin{array}{llcll}
 E_{b}+\Sigma_{1}&~~~~\beta&~~ 0 &\cdots & \cdots\\
  ~~~\beta& E_{b}-qV_{0}&~~\beta &~~ 0 & \cdots\\
 ~~~0 & \vdots &\ddots&\ddots&~~~~~~~~~\beta\\
 ~~~0 &\cdots& 0&~~\beta&E_{b}-(N-1)qV_{0}+ \Sigma_{N}
 \end{array}
 \right],
 \end{equation}
where $q$ is the average effective charge on each site. $G_{1N}$
can be obtained:
\begin{equation}
G_{1N}=\frac{(-1)^{N-1}\beta^{N-1}}{|zI-H_{eff}|},
\end{equation}
where $|zI-H_{eff}|$ denotes determinant. Then our task is to
deduce the expression of $|zI-H_{eff}|$. For convenience, we
define:
\begin{equation}
A_{n}=\left[
\begin{array}{llcll}
 a&~~~~\beta& 0 &\cdots & \cdots\\
  \beta& a-qV_{0}&~~\beta ~~& 0 & \cdots\\
 0 & \vdots &\ddots&\ddots&~~~~~~~\beta\\
 0 &\cdots& 0 &~~\beta&a-(n-1)qV_{0}
 \end{array}
 \right],
 \end{equation}
 and from linear algebra knowledge,
 \begin{equation}
 G_{1N}=\frac{(-1)^{N-1}\beta^{N-1}}{D_{1,N}-D_{1,N-1}\Sigma_{N}-D_{2,N}\Sigma_{1}+D_{2,N-1}\Sigma_{1}\Sigma_{N}}
 \end{equation}
where $a=E_{b}-E_{F}$, i.e. the site's energy measured from the
Fermi energy of the reservoirs without the electric field. $D_{PQ}
 (P=1, 2 ; Q=N-1, N)$ is the determinant of the matrix obtained from
A by taking rows and columns only in the range from P to Q
\cite{6}.

We use Generating Function method to obtain $D_{1N}$. Setting
$A_{n}=D_{1n}$, and $A_{N}=D_{1N}$, we find that $A_{n}$ satisfies
the following recursion relation:
\begin{equation}
A_{n}=[a-(n-1)qV_{0}]A_{n-1}-\beta^{2}A_{n-2}.
\end{equation}
If we define
\begin{equation}
 F(x)=\sum_{n=1}^{\infty}A_{n}x^{n},
\end{equation}
then we obtain that
\begin{equation}
 F(x)[1-ax+\beta^{2}x^{2}]+(qV_{0})x^{2}F'(x)=ax-\beta^{2}x^{2}.
\end{equation}
Since $qV_{0}$ is small, we can use a perturbation method, and
regard $qV_{0}$ as a perturbation parameter. Up to the first order
approximation, we have:
\begin{eqnarray}
F(x)&=&F_{0}(x)+(qV_{0})F_{1}(x),\\
F_{0}(x)&=&-1+\frac{1              }{1-ax+\beta^{2}x^{2}},\\
F_{1}(x)&=&\frac{-x^{2}F_{0}'(x)}{(1-ax+\beta^{2}x^{2})}.
\end{eqnarray}
Then we decompose $F(x)$ into a sum of several fractions, and
expand them to a power series, and obtain the $A_{n}$:
\begin{eqnarray}
\nonumber A_n &=&A_n^{(0)}+(qV_0)A_n^{(1)}, \\ \nonumber A_n^{(0)}
&=&\frac 1{2^{n+1}}\frac{{(a +\sqrt{a ^2-4\beta ^2})} ^{n+1}-{(a
-\sqrt{a^2-4\beta ^2})}^{n+1}}{{\sqrt{a^2-4\beta ^2}}}, \\
\nonumber A_n^{(1)} &=&-\frac{\beta ^{2n-1}} {8\mu
^3}{(n-1)(\lambda _1^n-\lambda _2^n)[1+n(1-\alpha ^2+\alpha \mu
\frac{\lambda _1^n+\lambda _2^n}{\lambda _1^n-\lambda _2^n})]}, \\
 \alpha  &=&\frac{a}{2\beta }, \\ \nonumber \mu
&=&\sqrt{\alpha ^2-1}, \\ \nonumber \lambda _1 &=&\frac{\alpha
-\sqrt{\alpha ^2-1}}\beta,  \\ \nonumber \lambda _2
&=&\frac{\alpha +\sqrt{\alpha ^2-1}}\beta.
\end{eqnarray}
In terms of our convention, $A_{N}$ is the determinant $D_{1,N}$.
Using similar steps, we can obtain $D_{1, N-1}$, $D_{2, N}$,
$D_{2, N-1}$, and then obtain $G_{1N}$. It is straightforward to
perform similar calculations for high order perturbation.

Then, we can calculate the current through the single molecular
dsDNA using wide energy band approximation. We assume
$\Delta_{K}({K=1,N}$) to be energy independent. For convenience,
we consider that the two electrodes have identical Fermi energies,
which are set to zero. For DNA molecules in equilibrium, the bases
of DNA are neutral, while in transport process, there are charges
introduced by the contacts, and charges in DNA will be
redistributed. Since the average charge $q$ on each site is small,
the derivation of the actual potential drop along DNA molecule
from the assumed linear drop can be ignored.

In previous efforts, many papers ignored the situation that the
sites' energies will vary along the molecular wire because of the
voltage drop in high-intensity electric field. Some papers did
consider this effect (e.g., Ref. \cite{7}), but they did not
obtain an analytic expression of the differential conductance. The
present paper achieves this task. In general, the modified
Hamiltonian is more complex. Since the Green's Function must be
obtained by calculating the inverse matrix, it is not a trivial
task, especially when the matrix is large.

Using the linear voltage drop approximation, we first obtain the
explicit expression of $G_{1N}$ up to the first order
perturbation, and get a more reasonable current-voltage relation.
We show the I-V curve in Fig.1 for a 4-nucleotide DNA, and Fig. 2
for a 30-nucleotide DNA. We find when considering the voltage drop
in the molecule, the current-voltage curves always become much
steeper near the threshold. This is in accordance with the
experiment of Porath {\it et al.} \cite{P}: some I-V curves in
their paper are much steeper near the threshold. Our result is
intuitional, because voltage drop effect makes the electron
transport more easily in strong electric field. Moreover, Fig. 3a
of Porath's paper contains two different I-V characteristics of
the same 30-base pair DNA sample. The difference between these two
curves may be the result of a sudden change in the DNA wire,
possibly a conformational change. In transport process, the charge
distribution along DNA may depend on DNA's chemical nature, for
example, the molecule's conformation. Although there was a similar
voltage drop in the sample of both cases, the observed
electrostatic drop effect are different. Finally, our work can be
used to calculate other molecular wires. Since the computational
complexity of our analytical deduction is not sensitive to the
number of sites, so we can calculate long chains.

An issue of interest is to investigate theoretically the transport
property of DNA molecules made of heterogeneous base sequences. We
are now working on this possibility.

In summary, we consider the electrostatic potential across the
DNA, which may change the site energy along the DNA with the
applied bias. This made the computation more difficult. We have
proposed an analytical computational method to deal with this
difficulty. It can be used to calculate the I-V characteristic of
homogeneous DNA wires of arbitrary length. We can use the
potential drop effect to explain recent single molecular DNA
experiments.

The authors are grateful to Guangping Gao for helpful discussions
and to Haijun Zhou for reading earlier versions of the manuscript.
We finally think Prof. J. Kong for modifying the manuscript.
%%%%%%%%%%%%%%%%%%%%%%%%%%%%%%%%%%%%%%%%%%%%%%%%%%%%%%%%%%%
%\begin{thebibliography}{}

%\end{thebibliography}
%%%%%%%%%%%%%%%%%%%%%%%%%%%%%%%%%%%%%%%%%%%%%%%%%%%%%%%%
\newpage
\begin{figure}[ht]
\scalebox{1.2}{\includegraphics{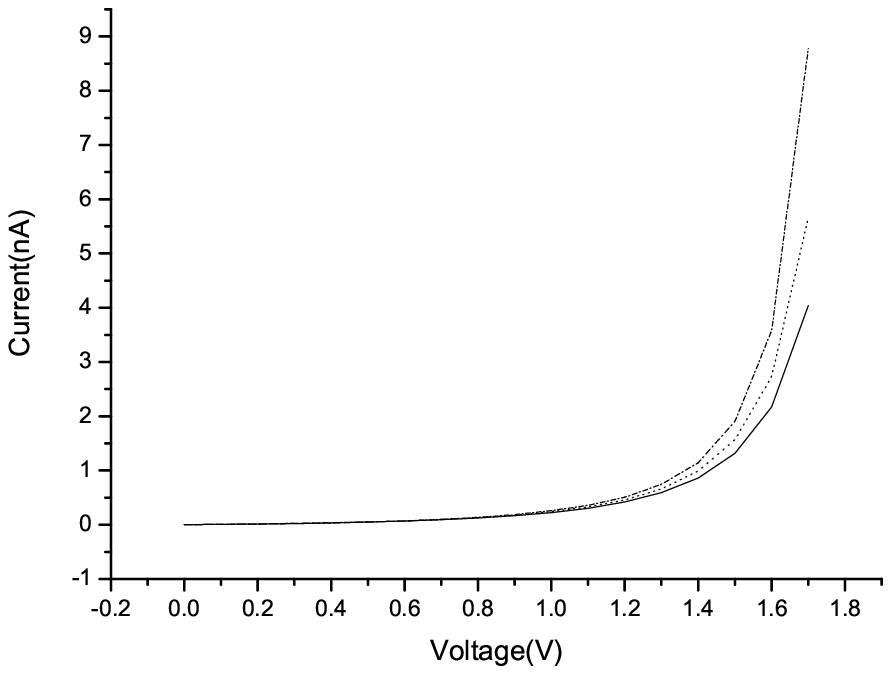}}
\caption{\label{fig1}Theoretically predicted current-voltage
characteristics of single molecular double stranded DNA.
Parameters: $\beta =-2.4 eV$, $V_{1}=V_{N}=0.5 eV$, $\gamma=10.0
eV$, $N=4$, and $E_{b}$ is $1.0 eV$ below $E_{F}$. The full line
is calculated with $q=0$, i.e., no site's energy shift effect. The
dashed curve is calculated with $q=0.05 e$. The dotted curve is
calculated with $q=0.1 e$. }
\end{figure}
%%%%%%%%%%%%%%%%%%%%%%%%%%%%%%%%%%%%%%%%%%%%%%%%%%%%%%%%
\newpage
\begin{figure}[ht]
\scalebox{1.2}{\includegraphics{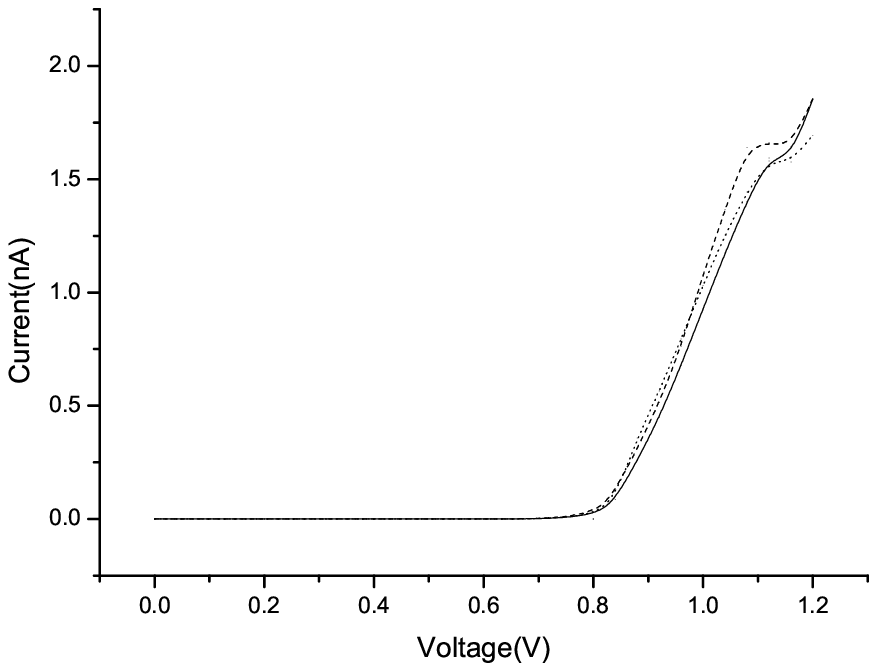}}
\caption{\label{fig2} Theoretically predicted current-voltage
characteristic of single molecular double stranded DNA.
Parameters: $\beta=-0.1 eV$, $V_{1}=V_{N}=3.0 eV$, $\gamma=10.0
eV$, $N=30$, and $E_{b}$ is $1.0 eV$ below $E_{F}$. The full line
is calculated with $q=0$, ie., no site's energy shift effect. The
dashed curve is calculated with $q=0.01e$. The dotted curve is
calculated with $q=0.05e$ .}
\end{figure}
\end{document}